\documentclass[final,sort&compress,a4paper,twocolumn]{article}
\usepackage{arxiv} 
\usepackage{authblk}
\usepackage[utf8]{inputenc}
\usepackage{amsmath}
\usepackage{amssymb}
\usepackage{graphicx}
\usepackage{acro}
\usepackage[hyphens]{url}
\usepackage[hidelinks]{hyperref}
\usepackage[capitalize]{cleveref}
\usepackage{enumitem}
\usepackage{multirow}
\usepackage{multicol}
\usepackage{siunitx}
\usepackage{float}
\usepackage[flushleft]{threeparttable}
\usepackage[table,xcdraw]{xcolor}
\usepackage[switch]{lineno}
\usepackage{array}
\usepackage{soul}
\usepackage{parskip}
\usepackage{caption}
\usepackage{subcaption}
\usepackage{tabularx}
\usepackage{placeins}
\usepackage{makecell}
\usepackage{tikz}
\usepackage[export]{adjustbox}
\usepackage{framed}
\usepackage{lipsum}
\usepackage[style=numeric-comp, sorting=none]{biblatex}
\usepackage[labelfont=bf, font={small}]{caption}
\usepackage{setspace}
\newcolumntype{L}[1]{>{\hsize=#1\hsize\raggedright\arraybackslash}X}%
\newcolumntype{R}[1]{>{\hsize=#1\hsize\raggedleft\arraybackslash}X}%
\newcolumntype{C}[1]{>{\hsize=#1\hsize\centering\arraybackslash}X}%

\fancypagestyle{firtpagestyleIEE}
{
    \fancyhf{}
    \fancyfoot[RO,LE]{\thepage}
    \fancyfoot[LO]{\begin{minipage}[t]{0.95\textwidth}\footnotesize Accepted in IEEE 20th International Symposium on Biomedical Imaging (ISBI) \copyright 2023 IEEE.  Personal use of this material is permitted.  Permission from IEEE must be obtained for all other uses, in any current or future media, including reprinting/republishing this material for advertising or promotional purposes, creating new collective works, for resale or redistribution to servers or lists, or reuse of any copyrighted component of this work in other works.\end{minipage}}
    
    \fancyhead{}
}

\begin{document}

\title{Estimating Head Motion from MR-Images} 

\author[1,*]{Clemens Pollak}
\author[1,*]{David Kügler}
\author[1,2,3]{Martin Reuter}

\affil[1]{\small{AI in Medical Imaging, German Center for Neurodegenerative Diseases (DZNE), Bonn, Germany}}
\affil[2]{\small{A.A. Martinos Center for Biomedical Imaging, Massachusetts General Hospital, Boston, MA, USA}}
\affil[3]{\small{Department of Radiology, Harvard Medical School, Boston, MA, USA}}



\date{}
\twocolumn[
\begin{@twocolumnfalse}
  \maketitle
  \vspace{-5ex}
  \begin{abstract}
  Head motion is an omnipresent confounder of magnetic resonance image (MRI) analyses as it systematically affects morphometric measurements, even when visual quality control is performed. In order to estimate subtle head motion, that remains undetected by experts, we introduce a deep learning method to predict in-scanner head motion directly from T1-weighted (T1w), T2-weighted (T2w) and fluid-attenuated inversion recovery (FLAIR) images using motion estimates from an in-scanner depth camera as ground truth. Since we work with data from compliant healthy participants of the Rhineland Study, head motion and resulting imaging artifacts are less prevalent than in most clinical cohorts and more difficult to detect. 
Our method demonstrates improved performance compared to state-of-the-art motion estimation methods and can quantify drift and respiration movement independently. Finally, on unseen data, our predictions preserve the known, significant correlation with age.
\end{abstract}

\thispagestyle{firtpagestyleIEE}

\vspace{4ex}


\vspace{2ex}

\noindent \textbf{\textit{Corresponding author}}: Martin Reuter (\texttt{martin.reuter[at]dzne.de})

\vspace{4ex}

\hrulefill

\vspace{8ex}

\end{@twocolumnfalse}
]
\section{Introduction}
\label{sec:intro}

\def\thefootnote{*}\footnotetext{These authors contributed equally to this work}

Head motion is a ubiquitous challenge for magnetic resonance image (MRI) acquisition. It causes a range of image artifacts that introduce bias in downstream analysis~\cite{reuter2015motion,alexander2016subtle,gilmore2021variations, savalia2017motion,pardoe2016motion,rosen2018quantitative},
which persists despite expert quality control~\cite{reuter2015motion, alexander2016subtle}.
While initially explored for clinical cohorts with increased motion levels~\cite{zaca2018method, zukic2022medical, pardoe2016motion, rosen2018quantitative} or induced motion~\cite{kustner2018automatic, reuter2015motion}, less research focuses on motion in studies of healthy, compliant population cohorts~\cite{fantini2018automatic, alexander2016subtle} such as the Rhineland Study~\cite{breteler2014mri, stocker2016big}. Critically, the lack of a sensitive and reliable motion estimation method to quantify subtle motion hinders the inclusion of motion estimates in statistical models to control motion-induced biases. 
For example, careful visual inspection of the Rhineland Study dataset, used in this paper, did not detect any cases with clearly visible motion artefacts that would warrant exclusion.  
Yet, even in the 75-participant subset reserved for testing, a statistically significant correlation of motion with age can be shown, underlining the need for sensitive estimation and control of head motion in MRI analyses. 

In this paper, we propose a method to directly estimate head motion from the acquired MR image. We measure head motion during MRI acquisition via head tracking with a depth camera and establish a ground truth motion score per sequence. This is contrary to the currently established paradigm of predicting discrete motion severity levels established via an expert manual quality control process~\cite{largent2021image, fantini2018automatic, kustner2018automatic, sujit2019automated, ma2020diagnostic, kustner2018automated, stpien2021fusion, zukic2022medical, lei2022artifact}. Expert ratings are limited by their subjectivity to the specific task and human perception~\cite{barrett2015task, prydeperformance, lei2022artifact} hindering their general utility, specifically for compliant, low-motion cohorts. Camera-based motion measurements, on the contrary, are objective and sensitive even for low-motion cohorts, where motion-induced image artifacts are almost invisible. 

Since the rise of deep learning, tools have been able to predict expert motion ratings with increasing accuracy~\cite{largent2021image, fantini2018automatic, kustner2018automatic, sujit2019automated, ma2020diagnostic, kustner2018automated, stpien2021fusion, zukic2022medical, lei2022artifact} sometimes addressing previous limitations, for example the subjectivity to the task~\cite{lei2022artifact}.
Currently, the only alternative to the prediction of expert labels is to predict a perceptual image similarity metric between low motion ''base-line'' images and high-motion images, which are retrospectively simulated~\cite{sciarra2022reference}.
This approach replaces the human annotation task by a comparison of images with a perceptual similarity metric (SSIM), which may also suffer from similar limitations as expert labels. Moreover, the methods accuracy on real-world data relies on realistic, high-quality simulation, which also has to be adapted to each acquisition sequence. Meanwhile, our method can be directly re-trained even on different modalities, without any changes. %
Recent work in the field of in-MRI motion tracking enabled highly accurate tracking of head-motion during acquisition with the MRI scanner~\cite{tisdall2012volumetric, andronesi2021motion}, optical cameras~\cite{slipsager2019markerless, pardoe2021estimation} or other devices~\cite{musa2022mri, laustsen2022tracking}. Yet, until tracking devices and methods are deployed to all imaging sites, image-derived motion estimates may help reduce potential motion induced biases -- even retrospectively. %

Our contributions are threefold -- we 1.~introduce -- for the first time -- the estimation of an objective motion score from images of three MRI sequences, 2.~present a deep-learning-based solution, which outperforms DenseNet and state-of-the-art quality/motion estimation methods on a dataset of compliant, low-motion participants, and 3.~quantify motion from respiration and (relaxation) drift. Finally, our method detects the significant, known correlation between predicted motion and age. We will publish our code on \href{https://github.com/Deep-MI/head-motion-from-MRI/}{GitHub}\footnote{\href{https://github.com/Deep-MI/head-motion-from-MRI/}{https://github.com/Deep-MI/head-motion-from-MRI/}}.

\section{Materials \& Methods}
\label{sec:method}

\begin{figure}[t]
    \centering
    \includegraphics[trim={0 .0cm 0 0.26},clip,width=3.2in]{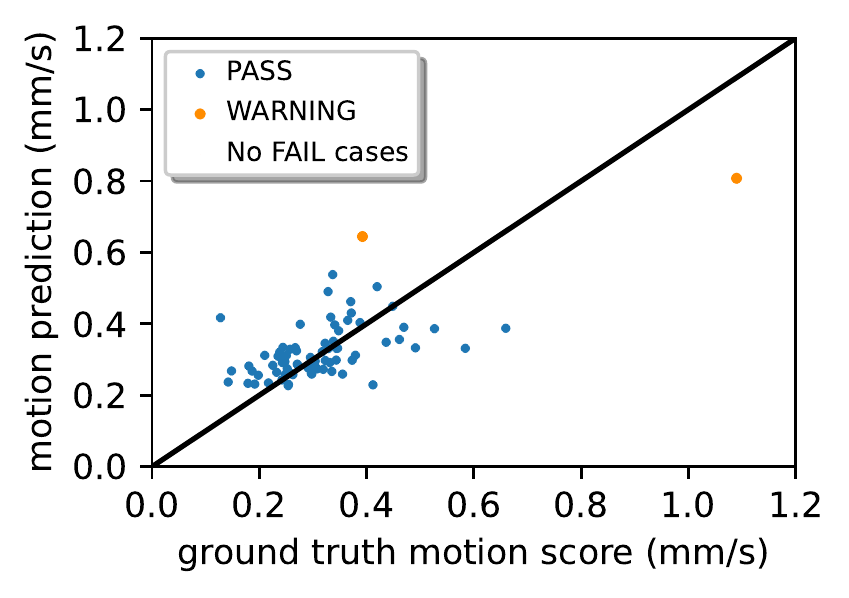}
    \caption{Plot of ground truth and predicted motion score. Blue \emph{PASS} and orange \emph{WARN} (visible artifacts) test images may be perfectly separated with a threshold of 0.6.}
    \label{fig:net_scatter}
\end{figure}

\subsection{Data acquisition}

The Rhineland study is an ongoing population study recruiting a representative cohort of healthy participants above the age of 30. Our dataset describes a subset (ages 30 to 95 years) of 500 participants (282 female) with T1w, T2w and FLAIR images (at 0.8, 0.8 and \SI{1.0}{\milli\meter} isotropic voxel size, respectively) following a standardized acquisition protocol\footnote{For details on the acquisition of 3D T1w MPRAGE (scan length 6.5 min), 3D T2w (4.6 min) and FLAIR (4.5 min) images, see Lohner et al.~\cite{lohner2022incidental}.}. %
This dataset also includes expert quality-labels of T1w images (\emph{PASS}, \emph{WARN} or \emph{FAIL}), where \emph{WARN} indicates visible and \emph{FAIL} strong artifacts (insufficient for downstream analysis). However, no images are rated as \emph{FAIL} and only 9 images as \emph{WARN}. 
The low number of \emph{WARN} and \emph{FAIL} cases is likely founded on high compliance and extensive efforts to reduce head motion during MRI acquisition including tight head padding, scheduled speaking breaks, careful participant instruction, and calming nature scenes shown during the scan.%

To quantify the participant's head motion, a video of depth images showing a portion of their face is collected concurrently to the scan~\cite{slipsager2019markerless}.
Individual frames are aligned with a reference frame resulting in time series of rigid transformations. To obtain a per-sequence, scalar \emph{motion score}, we 1.\ synchronize MRI and depth camera, 2.\ compute Jenkinson's transformation differences~\cite{jenkinson1999measuring} per pair of transformations, and 3.\ extract and average values for the duration of individual sequences. The Jenkinson's transformation difference summarizes rigid transformations by averaging displacements within a spherical head model. The final \emph{motion score} quantifies the average motion in millimeter per second. We randomly split the dataset into training, validation, and evaluation sets of 350, 75 and 75 participants, respectively.

\subsection{Distinguishing motion patterns}
\label{sec:freq}

In addition to the per-sequence motion score, we distinguish between three prominent types of in-scanner head motion corresponding to three frequency bands: %
i) head drift, ii) periodic motion due to breathing, and iii) ``noisy motion'', which we expect is hard to estimate.
To determine global filter thresholds, we estimate upper and lower median respiratory frequency of participants during the T1w sequence from an independent respiration sensor as \SI{0.1}{\hertz} and \SI{0.5}{\hertz}. We apply symmetric Butterworth filters (low-pass, band-pass and high-pass, respectively) to the time series of Jenkinson's transformation differences. From the filtered signals, we aggregate the motion score as before resulting in three frequency-dependent targets.

\subsection{Neural network \& training}

For the estimation of motion scores from 3D MR images, we adopt a fully convolutional neural network (CNN) from brain age estimation~\cite{peng2021accurate}, an established regression task in medical imaging. The lightweight architecture permits training the 3D CNN on a single NVIDIA A100 GPU with batch size two. Instead of directly regressing the motion score, we follow Peng et al.~\cite{peng2021accurate} in their approach and 1.\ generously define the expected range of motion [0, 3.12] mm/s, 2.\ split it into 40 bins of `prototypes', 3.\ for each prototype calculate the probability that the current motion score belongs into the prototype, and 4.\ train the CNN using a Kullback-Leibler loss (Adam optimizer for 500 epochs, approx.\ \SI{10}{\hour} on one A100 GPU). To reconstruct the motion score from the predicted probability distribution, we sum the product of prototype centers and predictions. %
Many standard data augmentation strategies are not suitable for motion estimation, since the re-sampling of images affects the image noise, which is why we avoid interpolation of images completely. Helpful data augmentation, on the other hand, include intensity scaling in the range [0.9,1.1] and random flipping with 30\% probability along all axes.
In our ablation study (Section~\ref{sec:ablation}), we explore different pre-processing operations. A focus on the 8 Least Significant Bits (LSB8) is useful for this task, leaving only an integer representation of the fine image differences. %

\begin{table}[t]
\centering
\begin{tabular}{| c | l | c | c |} 
     \hline
     & Method & R\textsuperscript{2} & Spr-$\rho$  \\
     \hline\hline
     \parbox[t]{2mm}{\multirow{4}{*}{\rotatebox[origin=c]{90}{re-trained}}} & Ours& \textbf{0.433} & \textbf{0.584} \\ 
     \cline{2-4}
     & DenseNet \cite{huang2017densely} & 0.395 & 0.447 \\
     \cline{2-4}
     & SFCN \cite{peng2021accurate} & 0.275 & 0.454 \\
     \cline{2-4}
     & MIQA CNN \cite{zukic2022medical} & -0.273 & 0.192 \\
     \hline\hline
     \parbox[t]{2mm}{\multirow{3}{*}{\rotatebox[origin=c]{90}{QCtools}}} & %
       MIQA MA / QS & - & 0.243 / 0.240\\
     \cline{2-4}
     & MIQA MA / QS\textsuperscript{1} & - & -\textsuperscript{2} / 0.338 \\
     \cline{2-4}
     & AES~\cite{zaca2018method} & - & 0.110 \\
     \hline
\end{tabular}

\caption{Our methods outperforms SOTA approaches in both Spearman's $\rho$ (Spr-$\rho$) and R\textsuperscript{2} scores (only valid for predicted motion scores) when predicting motion scores from T1w images on the test set. MA: motion artefact score, QS: quality score, \textsuperscript{1}images standardized, \textsuperscript{2}no motion detected  %
}
\label{tab:results}
\end{table}

\subsection{Evaluation \& statistical methods}

We evaluate the regression model with the coefficient of determination (R\textsuperscript{2} score -- a measure for the average error) and Spearman's rank correlation coefficient (Spearman's $\rho$ -- a measure for correct ranking). The R\textsuperscript{2} score normalizes the mean squared error to a range of $[-\infty, 1]$, where score \textless 0 indicates a prediction error worse than a constant prediction of the dataset mean and a score of 1 indicates perfect predictions. The Spearman's $\rho$, on the other hand, is defined in the range $[0,1]$. It is not affected by large, absolute errors of outliers and more sensitive to prediction errors, where the sampling of values is denser (i.e.\ more sensitive to small errors on values close together). We also analyze the rank correlation between motion and age using this method. We use the R\textsuperscript{2} score as the primary metric for ablation, and select parameters that have the highest R\textsuperscript{2} score on the validation set in experiments.

\section{Results}

We visualize the performance of our method on the unseen test set in Figure~\ref{fig:net_scatter}, which illustrates good correlation between ground truth measured motion (horizontal axis) and predictions from images (vertical). Perfect predictions would lie on the black line. Our method perfectly separates \emph{PASS} (no artifacts) and \emph{WARN} (mild artifacts) cases. A horizontal separation line at $\approx\SI{0.6}{\milli\meter/\second}$ can be found but no vertical line for the ground truth motion score.

\subsection{Comparison with state-of-the-art motion estimation}
\label{sec:results_general}

To the best of our knowledge, there is currently no competing method to predict measured, in-scanner head motion from MR images. Since quality estimation methods cannot be easily re-trained on our dataset, which has few \emph{WARN} and no \emph{FAIL} labels, we compare with the pre-trained MIQA quality estimator~\cite{zukic2022medical} and Average Edge Strength (AES)~\cite{zaca2018method}, a heuristic known to correlate with motion~\cite{zaca2018method}. Additionally, we compare our method with three deep learning architectures re-trained on our dataset: 
i) DenseNet~\cite{huang2017densely}, ii) SFCN~\cite{peng2021accurate}, a CNN for brain age prediction, and iii) the CNN used by MIQA~\cite{zukic2022medical}. Our method, which is an improvement of SFCN (e.g.\ initializer), achieves the best correlation with ground truth motion scores on the unseen test set (Table~\ref{tab:results}). The negative R\textsuperscript{2} of the re-trained MIQA CNN indicates failed generalization. %

The pre-trained MIQA tool aggregates predictions of expert ratings for 9 artifact types including a `motion artifacts' (MA) score into a continuous quality score (QS). We select their probabilities as potential correlates of our motion score.
Since MIQA has been trained on the \SI{1}{\milli\meter} T1w-PREDICT-HD dataset, we test, whether rescaling and resampling images with FastSurfer's~\cite{henschel2020fastsurfer} conform tool to \SI{1}{\milli\meter} reduces domain shift.
We measure a low but significant correlation between QS and motion score as well as between MA and motion score on the native \SI{0.8}{\milli\meter} images. While previous work reported that probabilities of ratings, like those of MIQA, quantify subtle differences in motion, despite binary ground truth labels~\cite{kustner2018automated}, MA probabilities are zero for all conformed images (consistent with the lack of manual \emph{FAIL} labels). QS, on the other hand, significantly correlates with the motion score.
In addition to the deep learning estimators, we evaluate the performance of AES~\cite{zaca2018method}, but do not find significant correlation between AES and the ground truth motion score.

\subsection{Generalization to T2/FLAIR and motion types}

\begin{table}[t]
    \centering
    \begin{tabular}{|l | c | c | c | c ||} 
        \hline
        Input & Target &R\textsuperscript{2} & Spr-$\rho$ \\
        \hline\hline
        T1 & motion score &0.433 & \textbf{0.584}\\
        \hline
        T2 & motion score & 0.362 & 0.556 \\
         \hline
        FLAIR & motion score & 0.299 & 0.489 \\
         \hline\hline
        T1 & drift & 0.183 & \textbf{0.637} \\
         \hline
        T1 & breathing band & 0.185 & 0.382 \\
         \hline
        T1 & noisy motion & 0.050 & 0.337 \\
         \hline

    \end{tabular}
    \caption{The evaluations show generalization of our method to predicting the motion score on T2w and FLAIR images and very promising Spr-$\rho$ performance for detection of drift on the test set.}
    \label{tab:generalize}
\end{table}

We test the generalizability of our method to new tasks by predicting the motion score for T2w and FLAIR images, as well as predicting drift and respiratory motion on T1w images. For each task we re-train the network architecture and show the results in Table~\ref{tab:generalize}. While the R2 score is not directly comparable across different tasks, we find good performance by purely re-training the model for these modalities. The Spearman's-$\rho$ in T2w and FLAIR experiments is similar to T1w experiments despite optimization on T1w only.%

To quantify different motion types, we define two distinct aggregates of participant motion: i) slow relaxation-drift over time, and ii) periodic head motion due to breathing. We filter motion estimates as described in Section~\ref{sec:freq} and train our architecture to predict the aggregates (Table~\ref{tab:generalize}). Our method can predict both slow drift (low frequency) motion and respiratory (medium frequency) motion from the T1w images. The aggregate of high frequencies, which are not associated with known motion types (noisy motion), cannot be predicted. %

\begin{table}[t]
\centering
\begin{tabular}{||l | c | c | c | c ||} 
     \hline
     Method & Images & R\textsuperscript{2} & Spr-$\rho$ \\ \hline\hline
     CNN MSE loss & T1 (LSB8) & 0.117 & 0.376 \\
     \hline\hline
     CNN ours & T1 (unprocessed) & 0.341 & 0.551 \\ \hline

    CNN ours & T1 (robust scaling) & 0.322 & 0.489 \\ \hline

    CNN ours & T1 (remove head) & 0.369 & 0.525 \\  \hline
    CNN ours & T1 (LSB8) & \textbf{0.393} & 0.491 \\ \hline

\end{tabular}
\caption{Ablation of loss function and image pre-processing on the validation set. LSB8: 8 Least Significant Bits. %
}
\label{tab:ablation}
\end{table}

\subsection{Ablation study}
\label{sec:ablation}

To optimize the parameters for our method, we compare choices for loss and preprocessing on the validation set. 
A critical finding of this work is that direct prediction trained using an Mean Squared Error (MSE) loss only achieves mild correlation with the motion score, while using a multi-dimensional probability distribution together with a Kullback-Leibler divergence loss yields a large performance uplift. For this MSE-loss ablation in Table~\ref{tab:ablation} (top), we reduce the last layer to a single output and remove the softmax.

Several measures of image quality have taken advantage of the background signal to determine a reduction in image quality~\cite{esteban2017mriqc, mortamet2009automatic}. %
Consequently, we explore the effects of four image pre-processing operations on the prediction quality in Table \ref{tab:ablation}: No pre-processing, FastSurfer's~\cite{henschel2020fastsurfer} robust scaling (removing 20\% of high intesity voxels), removing the head with FreeSurfer's head segmentation tool~\cite{fischl2012freesurfer, segonne2004hybrid} (the result is just the background) and dropping the Most Significant Bits leaving us with the 8 Least Significant Bits (LSB8). The latter approach, which drops the information about the absolute size of values directly on the images integer representation, surprisingly outperforms other ways of adjusting image intensities. %
This finding was validated in multiple experiments and across additional hypotheses.

\subsection{Correlation with age}

\begin{figure}[t]
    \centering
    \includegraphics[trim={0 .0cm 0 .26cm},clip,width=3.2in]{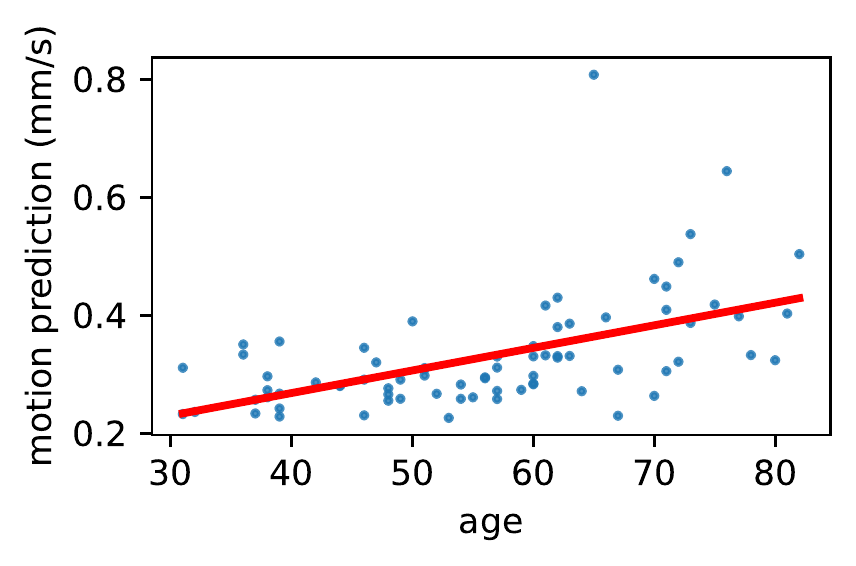}
    \caption{Age is significantly correlated with our motion predictions. This reflects the ground truth motion score labels (not shown) and is outlined by a linear fit, displayed as a red line.}
    \label{fig:age}
\end{figure}

In adult populations, increased head motion in the MR scanner is associated with the increased age of participants~\cite{madan2018age, savalia2017motion}. We can also measure this correlation within the ground truth motion score using Spearman's $\rho$ and find a significant correlation with age on the whole dataset as well as only on the test set (p \textless 0.001). Our predictions also present this correlation with age on the test set (p \textless 0.001) as illustrated in Figure~\ref{fig:age} with the linear fit. %

\section{Discussion \& conclusion}
We introduce a novel task of in-scanner head motion estimation directly from MR images. The presented deep learning method provides sensitive predictions of motion levels capturing subtle correlations with known confounders such as age -- all in unseen images that pass visual inspection.  
Two factors contribute to this sensitivity: predicting the probability distribution of `prototypes' together with the Kullback-Leibler divergence loss, and the input of least-significant-bit (LSB) images. Why LSB images are preferable to raw structural images or their background should be further investigated in future work. %

A well-known limitation of deep learning methods is the limited generalizability to unseen datasets. Large differences in motion levels between cohorts and the chosen acquisition parameters greatly affect the appearance of motion artifacts, hence we expect dedicated training datasets will be required to for re-training and generalization to unseen MR imaging sequences. 
Additionally, future work should investigate, whether motion estimation itself is also affected by biases, like age and diseases, which are known to be an indicator of increased motion levels.

Our method ranks images by their motion level better than comparable, state-of-the-art methods for MRI motion and quality estimation from expert ratings. Therefore, it may aid quality control procedures in the identification and exclusion of cases with artifacts, which is also indicated by the clear separation between \emph{PASS} and \emph{WARN} labels in our experiments (Figure~\ref{fig:net_scatter}). However, confirmation on a dataset with more strongly motion-affected cases is required. Additionally, our method transfers well to the prediction of alternative targets for respiratory- and drift motion and from T2w and FLAIR images with comparable accuracy. %

Finally, our method enables an analysis of other, perhaps unknown, correlates of motion as well as the integration of motion scores as a control variable in statistical models. This is particularly valuable in longitudinal cohort studies, like the Rhineland Study, to disentangle the bias of motion effects from other effects such as participant's age and diseases.

\section{Compliance with ethical standards}
\label{sec:ethics}

The Rhineland Study is carried out in accordance with the recommendations of the ICH‐GCP standards. Written informed consent was obtained from all participants in accordance with the Declaration of Helsinki. Approval was granted by the Ethics Committee of University Bonn. %

\section{Acknowledgments}
\label{sec:acknowledgments}

This work was supported by DZNE institutional funds, by the Federal Ministry of Education and Research of Germany (031L0206), and the Helmholtz-AI project DeGen (ZT-I-PF-5-078).
We thank the Rhineland Study group (PI Monique Breteler) for supporting the data acquisition and management.
The authors do not have any conflict of interest.

\printbibliography

\setlength{\FrameRule}{0pt}
\setcounter{table}{0}
\setcounter{figure}{0}
\renewcommand{\thesubsection}{\Alph{subsection}}
\renewcommand\theequation{A.\arabic{equation}}
\renewcommand\thetable{A\arabic{table}}
\renewcommand\thefigure{A\arabic{figure}}

\end{document}